\documentclass[12pt,aps,prd,abstract]{revtex4-2}   	
\usepackage{mwe,float}
\usepackage{caption,subcaption, url}
\usepackage{natbib}
\usepackage[nointegrals]{wasysym}
\usepackage{rotating}
\usepackage{graphicx,bm}
\usepackage{floatpag} 
\rotfloatpagestyle{empty}
\usepackage{amsmath}
\usepackage{amsthm}
\usepackage{amssymb}
\usepackage{latexsym}
\usepackage{hyperref,rotating}
\usepackage{xpatch,bigints,slashed}
\usepackage{gensymb}
\usepackage{mathdots}

\newcommand{\qand}{\quad \mbox{and} \quad}
\newcommand{\qor}{\quad \mbox{or} \quad}

\usepackage{kecpm,revsymb4-2}

\newcommand{\bex}{\begin{example}}
\newcommand{\eex}{\end{example}}
\newcommand{\besp}{\begin{split}}
\newcommand{\ensp}{\end{split}}

\renewcommand{\thefootnote}{\fnsymbol{footnote}}

\newcommand{\by}{\times}

\newcommand{\ovl}{\overline}

\newcommand{\btab}{\begin{tabular}}
\newcommand{\etab}{\end{tabular}}
\newcommand{\barr}{\begin{array}}
\newcommand{\earr}{\end{array}}
\newcommand{\bpm}{\begin{pmatrix}}
\newcommand{\epm}{\end{pmatrix}}
\newcommand{\bit}{\begin{itemize}}
\newcommand{\eit}{\end{itemize}}
\newcommand{\ben}{\begin{enumerate}}
\newcommand{\een}{\end{enumerate}}
\newcommand{\bct}{\begin{center}}
\newcommand{\ect}{\end{center}}

\newcommand{\ra}{\rangle}
\newcommand{\la}{\langle}
\newcommand{\bes}{\begin{split}}
\newcommand{\ens}{\end{split}}

\newcommand{\lt}{\left}
\newcommand{\rt}{\right}

\begin{document}
\title{Zero-point energies, dark matter, and dark energy}

\author{Kevin Cahill}
\affiliation{Department of Physics and Astronomy\\
University of New Mexico\\
Albuquerque, New Mexico 87106}
\date{\today}

\begin{abstract}
A quantum field theory has
finite zero-point energy if 
the sum over all 
boson modes $b$ 
of the $n$th power of
the boson mass 
$ m_b^n $ equals
the sum 
over all fermion modes $f$ 
of the $n$th power of
the fermion mass $ m_f^n $
for $n= 0$, 2, and 4.
The zero-point energy
of a theory that satisfies
these three conditions with  
otherwise random masses
is huge compared
to the density of dark energy.
But if in addition to satisfying
these conditions, the sum of
$m_b^4 \log m_b/\mu$ 
over all 
boson modes $b$ equals
the sum of $ m_f^4 \log m_f/\mu $
over all fermion modes $f$,
then the zero-point energy of the
theory is zero.
The value of the mass parameter 
$\mu$ is irrelevant in view of
the third condition ($n=4$).
\par
The particles of the standard model
do not remotely obey 
any of these four conditions.
But an inclusive theory that describes
the particles of the standard model,
the particles of dark matter,
and all particles that have not
yet been detected 
might satisfy all four conditions
if pseudomasses are associated
with the
mean values in the vacuum
of the divergences of the
interactions of the inclusive model.
Dark energy then would be 
the finite potential energy 
of the inclusive theory.
\end{abstract}

\maketitle

\section{Introduction
\label{Introduction sec}}

The vacuum energy
of a generic, noninteracting,
exactly soluble
quantum field theory
is quartically divergent.
But theories in which the sum 
of $m_b^n$ over all boson modes $b$
equals the sum of $m_f^n$ over
all fermion modes $f$ for
$n =0$ and $n=2$
\begin{equation}
\sum_b m_b^n ={} \sum_f m_f^n
\quad \mbox{for} \quad n = 0, 2
\label{the quadratic sum}
\end{equation}
have zero-point energies that
are at most logarithmically divergent.
In these sums over modes, 
the Higgs occurs once,
the photon twice, the $W^+$ three times, 
and the electron-positron field four times.
Theories with suitably
broken supersymmetry
satisfy these 
conditions~\cite{Fayet:1978rb,Ferrara:1979wa},
but the particles of the standard model
with (or without) the graviton do not:
\begin{equation}
\sum_b m_b^0 {} - \sum_f m_f^0 ={} -66
\qand
\sum_b m_b^2 {} - \sum_f m_f^2 ={}
- 280,010 \>\, \mbox{GeV}^2.
\label{standard-model particles don't}
\end{equation}
\par
It is shown
in section~\ref{Zero-point energies that cancel sec}
that if the three conditions
\begin{equation}
\sum_b m_b^n ={} \sum_f m_f^n
\quad \mbox{for} \quad 
n = 0, \, 2, \, \mbox{and} \>\, 4
\label{the three conditions}
\end{equation}
are satisfied, then the 
zero-point energy of the theory
is finite.
It also is shown in 
section~\ref{Zero-point energies that cancel sec} 
that if a theory obeys these three
conditions (\ref{the three conditions})
as well as the fourth condition
\begin{equation}
\sum_b m_b^4 \, \log \frac{m_b}{\mu}
={} \sum_f m_f^4 \, \log \frac{m_f}{\mu}
\label{the fourth condition}
\end{equation}
then its zero-point energy
vanishes.
This fourth condition is independent 
of the value of the mass parameter $\mu$
when the boson and fermion masses
obey the third condition 
(\ref{the three conditions})
for $n=4$\@.
I do not know of a symmetry principle
that implies the third condition
(\ref{the three conditions}, $n=4$),
let alone the fourth condition
(\ref{the fourth condition})\@.

\par
The numerical examples
of section~\ref{Numerical Examples sec}
show that theories whose
boson and fermion masses
obey the three conditions
(\ref{the three conditions})
with otherwise random masses
up to 10 GeV/c$^2$
have zero-point energy densities
that while finite are
42 orders of magnitude
greater than the
density of dark energy
$\rho_{\mbox{\scriptsize de}} ={} 
(2.25 \by 10^{-3} \, \mbox{eV} )^4
={} 5.32 \times 10^{-10}\) 
J\,m$^{-3}$~\citep{Aghanim:2018eyx}\@.
Presumably a
symmetry principle makes the
zero-point energy density $\rho_0$ 
tiny or makes the masses
obey the fourth condition
(\ref{the fourth condition}) so that 
$\rho_0$ vanishes.
\par
In most theories, the energy density
of the vacuum is due to interactions
as well as to zero-point energies.
Section~\ref{Vacuum energy sec}
relates the divergences of the interactions
to pseudomass terms and
discusses the energy density
of the vacuum the way 
section~\ref{Zero-point energies that cancel sec}
discusses zero-point energy densities.
It is suggested in 
section~\ref{Inclusive theories sec}
that ordinary matter,
dark matter, and dark energy
are aspects of 
a single inclusive theory
whose masses and pseudomasses
satisfy four conditions that 
generalize the four conditions
(\ref{the three conditions} and
\ref{the fourth condition})\@.
In such an inclusive theory,
the energy density of the vacuum
would be finite and due 
interactions and not to 
zero-point energies.
Dark energy  
would be the potential 
energy of the inclusive theory,
a kind of 
quintessence~\cite{Ratra:1987rm,Caldwell:1997ii}.

\section{Zero-point energies that cancel
\label{Zero-point energies that cancel sec}}
The zero-point energy density of a
bosonic mode of mass \(m\) is 
the $K \to \infty$ limit of the integral
\begin{equation}
\frac{1}{4\pi^2}
\int_0^K k^2 \sqrt{k^2 + m^2} \, dk
={}
\frac{K}{32\pi^2} \left(2 K^2+m^2\right) \sqrt{K^2+m^2} -
\frac{m^4}{32\pi^2} \log \left(
\sqrt{1 + \frac{K^2}{m^2}}
+ \frac{K}{m} \right).
\label {An integral}
\end{equation}
The zero-point energy density of a
fermion mode is the same integral
apart from an overall minus sign.
Zero-point energies cancel
between bosons and fermions of the
same mass.
\par
Let us consider a theory
that obeys the three
conditions (\ref{the three conditions}):
\begin{enumerate}
\item
The number of boson modes
is the same as the
number of fermion modes
\begin{equation}
\sum_b m_b^0 ={} \sum_f m_f^0.
\label{condition one}
\end{equation}
\item
The sum of the squares of the masses
of the bosons is the same as 
sum of the squares of the masses
of the fermions
\begin{equation}
\sum_b m_b^2 ={} \sum_f m_f^2.
\label{condition two}
\end{equation}
\item
The sum of the fourth power 
of the masses
of the bosons is the same as 
the sum of the fourth power 
of the masses
of the fermions
\begin{equation}
\sum_b m_b^4 ={} \sum_f m_f^4.
\label{condition three}
\end{equation}
\end{enumerate}
\par
Rewriting the logarithm of the
integral (\ref{An integral}), 
we have
\begin{equation}
\begin{split}
\frac{1}{4\pi^2}
\int_0^K k^2 \sqrt{k^2 + m^2} \, dk
={}&
\frac{2 K^3+m^2K}{32\pi^2} 
\sqrt{K^2+m^2} -
\frac{m^4}{32\pi^2}
\log \left[\frac{2K}{m}
\lt(\half + \sqrt{\fourth + \frac{m^2}{4K^2}}
\rt)\right] .
\label {An integral with mu}
\end{split}
\end{equation}
The expansion of this integral
in powers of \(m/K\) is
\begin{equation}
\begin{split}
\frac{1}{4\pi^2}
\int_0^K k^2 \sqrt{k^2 + m^2} \, dk
={}&
\frac{K^4 + m^2 K^2}{16\pi^2} -
\frac{m^4}{32\pi^2} \log\frac{K}{\mu}
+ \frac{m^4}{128\pi^2}
\lt( 1 + 4\log \frac{m}{2\mu} \rt) + \dots
\label{expansion of this integral}
\end{split}
\end{equation}
in which $\mu$ is an arbitrary
energy, and the dots indicate terms
proportional to positive powers
of \( m/K \) which vanish as $K \to \infty$\@.
The quartically divergent term 
\( K^4/16 \pi^2 \) cancels in theories
that satisfy the first condition,
that the number
of boson modes is equal to the number
of fermion modes.
The quadratically divergent term 
\( m^2 K^2 / 16 \pi^2 \)
cancels in theories that obey the 
second condition, that the sum 
of the squared masses of
the bosons is the same as 
the sum of the squared masses of the fermions.
The logarithmically divergent term 
\({}- (m^4/32\pi^2) \, \log(K/\mu) \) cancels
when the sum of the fourth power of the  masses of
the bosons is the same as that of the fermions.
The zero-point energy density 
$\rho_0$ of a theory 
that obeys the three conditions 
(\ref{condition one}, \ref{condition two},
and \ref{condition three}) 
is~\cite{Kamenshchik:2018ttr} 
\begin{equation}
\rho_0 ={} \frac{1}{32 \pi^2}
\lt[
\sum_b m_b^4 \log \lt( \frac{m_b}{2\mu} \rt)
- 
\sum_f m_f^4 \log \lt( \frac{m_f}{2\mu} \rt) 
\rt] 
\label {zero-point energy}
\end{equation}
which  is independent
of the arbitrary energy scale \(2\mu\)
because of condition 3\@.
But if the masses 
of a theory obey 
not only the three conditions 
(\ref{condition one}, \ref{condition two},
and \ref{condition three})
but also the fourth condition 
(\ref{the fourth condition})
\begin{equation}
\sum_b m_b^4 \, \log \frac{m_b}{2\mu}
={} \sum_f m_f^4 \, \log \frac{m_f}{2\mu}
\label{condition four}
\end{equation}
then the zero-point energy
density of the theory vanishes, $\rho_0=0$\@.

\section{Numerical Examples
\label{Numerical Examples sec}}

A single Majorana fermion
of mass \(m\)
has two modes, spin up
and spin down.
The first condition (\ref{condition one}) 
requires two spin-zero bosons.
The second and third conditions 
(\ref{condition two} and \ref{condition three})
then are 
\( 2m^2 = \mu_1^2 + \mu_2^2 \)
and \( 2m^4 = \mu_1^4 + \mu_2^4 \)\@.
The only solution is
\( \mu_1 = \mu_2 = m \)\@. 
The zero-point energy density
(\ref{zero-point energy})
of this theory vanishes,  \(\rho_0 = 0\)\@.
\par
A more interesting case is
that of two Majorana fermions,
one of mass \(m_1\) and
the one of mass \(m_2\)\@.
Canceling the \(K^4\)
divergence of the integral
(\ref{expansion of this integral}) 
requires four bosonic modes,
for instance, four spinless bosons
of masses \( \mu_1, \mu_2, \mu_3, \)
and \( \mu_4\)\@.
One may cancel the 
remaining divergences 
of the integral (\ref{expansion of this integral})
by imposing the second and third conditions 
(\ref{condition two} and \ref{condition three}):
\begin{eqnarray}
2m_1^2 + 2 m_2^2 ={}&
\mu_1^2 + \mu_2^2 + \mu_3^2
+ \mu_4^2
\label{condition deux}
\\
2m_1^4 + 2 m_2^4 ={}&
\mu_1^4 + \mu_2^4 + \mu_3^4
+ \mu_4^4.
\label{condition trois}
\end{eqnarray}
One sets 
\begin{equation}
\a ={} 2m_1^2 + 2 m_2^2
- \mu_1^2 - \mu_2^2
\qand
\b ={} 2m_1^4 + 2 m_2^4
- \mu_1^4 - \mu_2^4 .
\end{equation}
Both must be positive,
$ \a \ge 0$ and $\b \ge 0$,
because masses
$\mu_3$ and $\mu_4$ are real.
We can satisfy version
(\ref{condition deux})
of the second condition 
(\ref{condition two}) by setting
$ \mu_4^2 ={} \a - \mu_3^2 $,
which must be positive,
\( \mu_4^2 \ge 0 \)\@.
Version (\ref{condition trois})
of the third condition 
(\ref{condition three})
then requires that
$ \b = \mu_3^4 + \mu_4^4
=  \mu_3^4 + ( \a - \mu_3^2 )^2$
which is a quadratic equation
for \(\mu_3^2\)
\begin{equation}
2 \mu_3^4 - 2 \a \mu_3^2 + \a^2 - \b ={} 0.
\label{quadratic equation for mu32}
\end{equation}
Its solutions are
\begin{equation}
\mu_3^2 = \frac{1}{4}
\lt( 2 \a \pm \sqrt{4 \a^2 - 8(\a^2 - \b)}
\rt)
= \half \lt( \a \pm \sqrt{2 \b - \a^2} \rt) .
\label{mu3 squared}
\end{equation}
The equation 
$ \mu_4^2 ={} \a - \mu_3^2 $
which implements the second condition 
(\ref{condition two})
then says that
\begin{equation}
\mu_4^2 ={} \a - 
\half \lt( \a \pm \sqrt{2 \b - \a^2} \rt)
= \half \lt( \a \mp \sqrt{2 \b - \a^2} \rt) .
\label{mu4 squared}
\end{equation}
So if $ \a \ge 0$, $\b \ge 0$,
and $\b \le \a^2 \le 2\b$, then
$\mu_3^2$ and \( \mu_4^2 \) 
are positive,
and the zero-point energy density
(\ref{zero-point energy})
of the theory is
\begin{equation}
\rho_0 ={} \frac{1}{32 \pi^2} \lt\{\lt[
\sum_{b=1}^4 m_b^4 \log \lt( \frac{m_b}{2\mu} \rt) \rt]
- 2 m_1^4 \log \lt( \frac{m_1}{2\mu} \rt)
- 2 m_2^4 \log \lt( \frac{m_2}{2\mu} \rt)
\rt\} .
\label {zero-point energy for 2 fermions}
\end{equation}
\par
\begin{figure}
\centering
\includegraphics[width=4in]{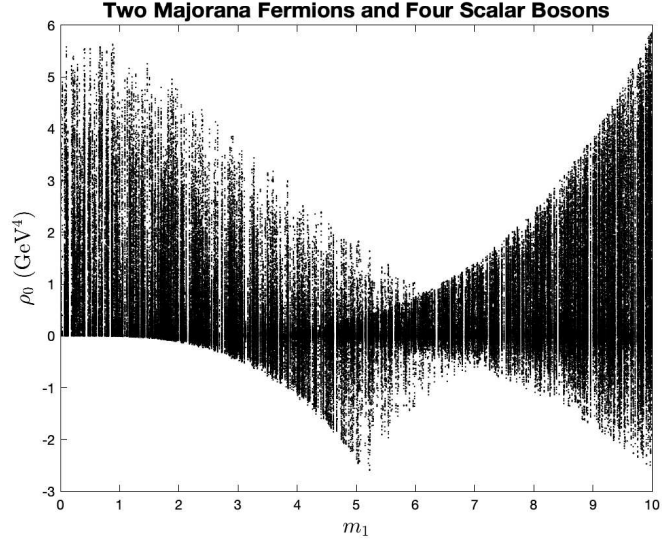}
\caption{A plot of the 234,340 values
of zero-point energy density 
$\rho_0$ (equation \ref{zero-point energy})
that satisfy the three conditions
(\ref{the three conditions}) for $10^6$ 
random values of the fermion
masses $m_1$ and $m_2$ 
and of the boson masses $\mu_1$ and $\mu_2$
between 0 and 10 GeV\@.}
\label {exfig1 fig}
\end{figure}
\begin{figure}
\centering
\includegraphics[width=4in]{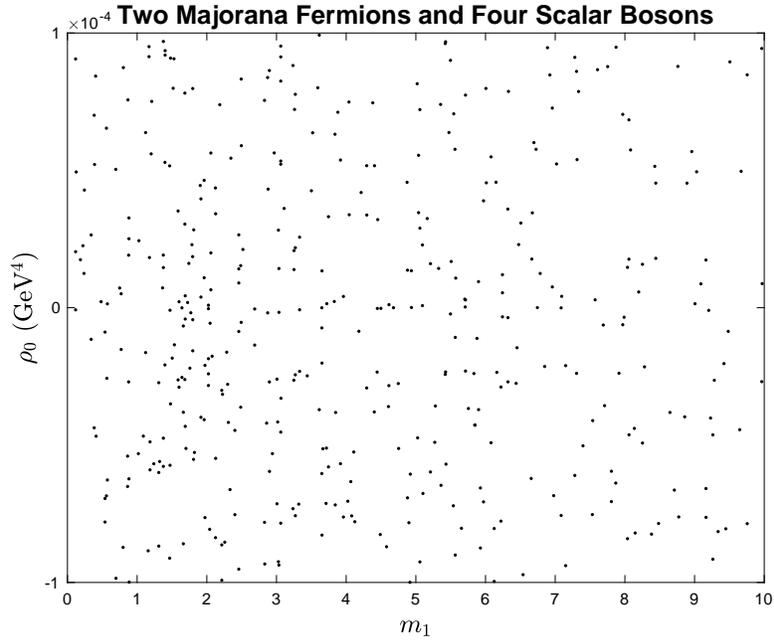}
\caption{A plot of the 421 values
(\ref{zero-point energy}) 
of the zero-point energy density
$\rho_0$
that lie in the interval
$-10^{-4} < \rho_0 < 10^{-4}$ GeV$^4$
for $10^6$ 
random values 
between 0 and 10 GeV
of the fermion
masses $m_1$ and $m_2$ 
and of the boson masses 
$\mu_1$ and $\mu_2$\@.}
\label {exfig2 fig}
\end{figure}

Figure~\ref{exfig1 fig} displays the
234,340 values of this energy density
that were positive
for $10^6$ random values
of these fermion masses
$m_1$ and $m_2$ 
and of the boson masses 
$\mu_1$ and $\mu_2$
all within the interval (0,10) GeV\@.
About 23\% of the random masses
give positive values for
$\mu_3^2$ and $\mu_4^2$\@.
Small values of $\rho_0$
are possible as shown 
in Fig.~\ref{exfig2 fig}, but
only 421 values were
within the range
$-10^{-4} < \rho_0 
< 10^{-4} \, \mbox{GeV}^4$
and only 5 were within the range
$-10^{-8} < \rho_0 
< 10^{-8} \, \mbox{GeV}^4$\@.
The density $10^{-4}$
GeV$^4$ is greater than the
density of dark energy
$\rho_{\mbox{\scriptsize de}} = 
(2.25 \by 10^{-3} \, \mbox{eV} )^4$
by $ 3.9 \by 10^{42}$;
the density $10^{-8}$
GeV$^4$ is greater than 
$\rho_{\mbox{\scriptsize de}}$
by $ 3.9 \by 10^{38}$\@.

\par
The algebra of the second example also
applies to a theory with $2n$ Majorana
fermions (or $n$ Dirac fermions)
and $4n$ scalar bosons.
Let $s_2$ and $s_4$ be the sums of the
second and fourth powers of the
fermion masses
\begin{equation}
s_2 ={} 2 \sum_{i=1}^{2n} m_i^2
\qand
s_4 ={} 2 \sum_{i=1}^{2n} m_i^4 
\end{equation}
in which the factor of 2 counts the
two spin states of the Majorana
fermions.
Let $\s_2$
and $\s_4$ be the sums of all but two
of the second and fourth powers of the 
boson masses
\begin{equation}
\s_2 = \sum_{i=1}^{4n-2}
\mu_i^2
\qand
\s_4 = \sum_{i=1}^{4n-2}
\mu_i^4 .
\end{equation}
We set
$\a ={} s_2 - \s_2$
and
$\b ={} s_4 -\s_4$.
Both must be positive,
$ \a \ge 0$ and $\b \ge 0$\@.
Then \(\mu_{4n-1}^2\) must satisfy
the analog of the quadratic equation 
(\ref{quadratic equation for mu32})
\begin{equation}
2 \mu_{4n-1}^4 - 2 \a \mu_{4n-1}^2 + \a^2 - \b ={} 0 .
\end{equation}
Like
(\ref{mu3 squared}) and (\ref{mu4 squared}),
the solutions are 
\begin{equation}
\begin{split}
\mu_{4n-1}^2 = {}&
\half \lt( \a \pm \sqrt{2 \b - \a^2} \rt)
\\
\mu_{4n}^2 = {}& 
 \half \lt( \a \mp \sqrt{2 \b - \a^2} \rt),
\label{mu4n-1 squared}
\end{split}
\end{equation}
and $\mu_{4n-1}^2$ and $\mu_{4n}^2$
will be positive if
$ \b < \a^2 < 2\b $\@.

\par
\begin{figure}
\centering
\includegraphics[width=4in]{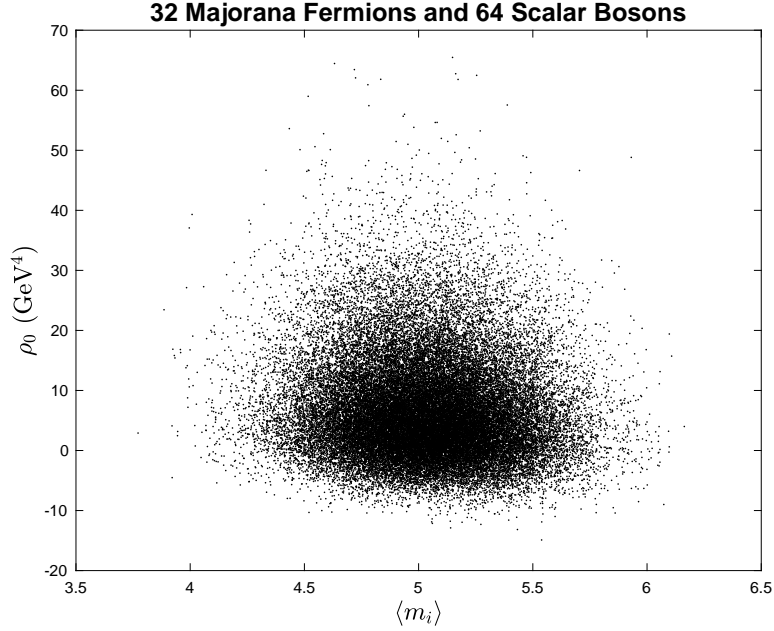}
\caption{A plot of the zero-point energy
(\ref{zero-point energy}) for
random values of 32 fermion
masses $m_i$ and 64 boson masses 
$\mu_i$ all within the interval
$0 < m_i, \mu_i < 10$\@.}
\label {exfig3 fig}
\end{figure}

Figure (\ref{exfig3 fig}) plots the
zero-point energy density $\rho_0$ 
(\ref{zero-point energy}) 
for $10^6$ random 
sets of 32 Majorana fermion masses
and 64 scalar boson masses
all within the interval (0,10) GeV
against the average $\la m_i \ra$
of the 32 fermion masses of each set.
About 5\% or 48,311 random sets
of masses
survived the  cuts $\mu_{63}^2 > 0$
and $\mu_{64}^2 > 0$,
and only 5 sets gave densities in
the range $-10^{-4} < \rho_0 < 10^{-4}$ 
GeV$^4$\@.
The average of the average $\la m_i \ra$
was $\ovl{\la m_i\ra} = 5.03$ GeV\@.
The average value of the  
zero-point energy density (\ref{zero-point energy}) 
was $ \la \rho_0 \ra ={} 6.83 \, \mbox{GeV}^4$,
which exceeds the density 
$\rho_{\mbox{\scriptsize de}}$ by 
more than 47
orders of magnitude.

\par
\begin{figure}
\centering
\includegraphics[width=4in]{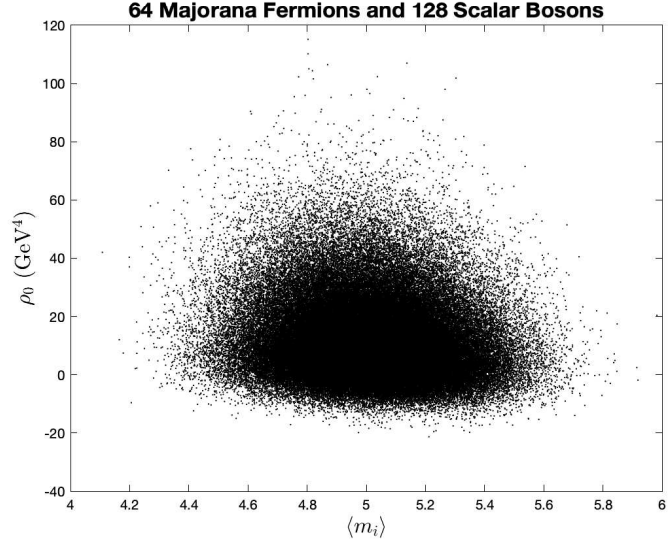}
\caption{A plot of the zero-point energy
(\ref{zero-point energy}) for
random values of 64 fermion
masses $m_i$ and 128 boson masses 
$\mu_i$ all within the interval
$0 < m_i, \mu_i < 10$\@.}
\label {exfig4 fig}
\end{figure}

Figure (\ref{exfig4 fig}) plots the
zero-point energy density $\rho_0$ 
(\ref{zero-point energy})
for $4 \by 10^6$ random 
sets of 64 Majorana fermion masses
and 128 scalar boson masses
all within the interval (0,10) GeV
against the average $\la m_i \ra$
of the 64 fermion masses of each set.
About 3.5\% or 140,601 
random sets of masses
survived the cuts $\mu_{127}^2 > 0$
and $\mu_{128}^2 > 0$\@.
The average of the average $\la m_i \ra$
is $\ovl{\la m_i\ra} = 5.01$ GeV\@.
The average value of the  
zero-point energy density (\ref{zero-point energy}) 
is $ \la \rho_0 \ra ={} 
11.78 \, \mbox{GeV}^4$,
which exceeds the density 
$\rho_{\mbox{\scriptsize de}}$ by 
47 orders of magnitude.

\par
\begin{figure}
\centering
\includegraphics[width=4in]{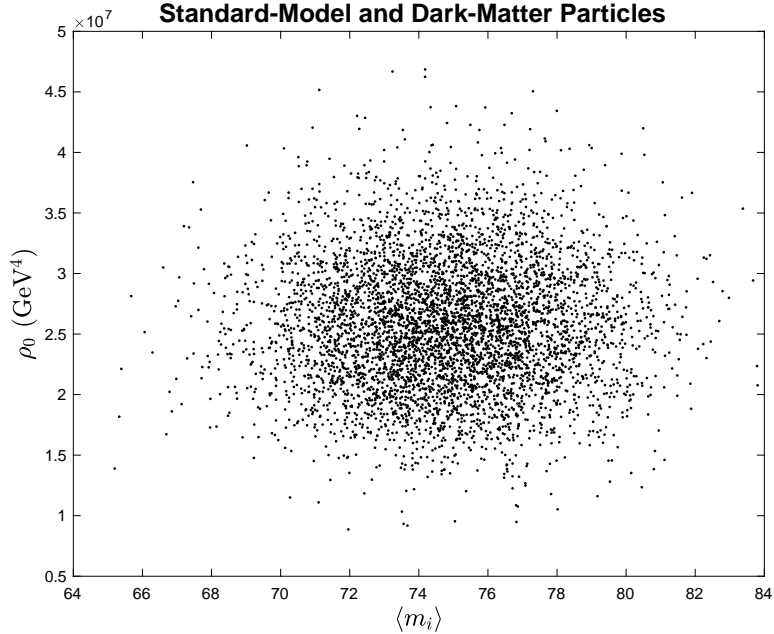}
\caption{A plot of the zero-point energy
(\ref{zero-point energy}) for
the masses of the
particles of the standard model 
and for values of 
the masses of the dark-matter
particles distributed uniformly and
randomly  on the interval
$0 < m_i, \mu_i < 200$\@.
In all 100 Majorana fermions
and 200 bosonic modes were used.}
\label {exfig5 fig}
\end{figure}

The fifth example has 100 fermion
masses and 200 boson masses, 
some of which are the actual
quark, lepton,
and gauge-boson masses.
The quark masses are $m_u = 2.2$,
$m_d = 4.7$, $m_s = 95$, $m_c = 1275$,
$m_b = 4180$, and 
$m_t = 173,000$ MeV\@.
The charged lepton masses are
$m_e = 0.511$, $m_\mu = 105.66$,
and $m_\tau = 1776.9$ MeV\@.
The neutrino masses are known to 
be less than 2 eV, probably less than
0.19 eV, and the Planck collaboration
gives the sum of their masses as less
than 0.23 eV\@.  
I took them to be three Dirac
fermions of mass 0.05 eV\@.
The photon and graviton are massless,
and each has 2 modes. 
Each heavy gauge boson
has 3 modes with
$m_W = 80,379$ and $m_Z = 91,187.6$ MeV\@.
The Higgs has mass 125,180 MeV\@.
I ran $10^8$ sets of masses that 
included the physical masses of
the known particles and random values
of the masses of dark-matter particles
uniformly distributed on the interval
$(0,200)$ GeV\@.
Of these $10^8$ sets of masses, 
5,102 survived the cuts;
their values of $\rho_0$ are plotted in
Fig.~\ref{exfig5 fig} against 
the average $\la m_i \ra$
of the 100 fermion masses.
The mean value $\ovl{\la m_i \ra}$
of the average fermion mass is
74.78 GeV\@.
The average zero-point energy density 
(\ref{zero-point energy}) was
$\la \rho_0 \ra ={} 2.6 \by 10^7 
\> \mbox{GeV}^4$,
which is 54 orders of magnitude
greater than the density of dark energy.
The minimum energy density was 
$8.9 \by 10^6$ GeV$^4$\@.

\section{Vacuum energy}
\label{Vacuum energy sec}

The interactions of a typical
quantum field theory cause
the vacuum energy density
to diverge logarithmically
or quadratically but not quartically. 
In a theory whose interactions
result in a quadratically divergent
vacuum energy density
${} \pm E_{\mbox{\scriptsize quad}}^2 \, K^2$
(as $K \to \infty$),
one may use the relation
$ {}   
m_{\mbox{\scriptsize quad}}^2K^2/(16 \pi^2) 
= {} E_{\mbox{\scriptsize quad}}^2 \, K^2$
to associate a quadratic pseudomass
$m_{\mbox{\scriptsize quad}} 
={} 4 \pi E_{\mbox{\scriptsize quad}}$
with a fictitious boson if the
vacuum energy is positive
or with a fictitious fermion if the
vacuum energy is negative.
One then would generalize the
second condition 
(\ref{condition two}) to
\begin{equation}
m_{\mbox{\scriptsize quad}}^2
+ \sum_b m_b^2 
={} \sum_f m_f^2
\qor
\sum_b m_b^2 
={} m_{\mbox{\scriptsize quad}}^2
+ \sum_f m_f^2.
\label{pseudo condition two}
\end{equation}
%sections~\ref{Zero-point energies that cancel sec}
%and \ref{Numerical Examples sec}.
If the interactions of a theory 
result in a 
vacuum energy density
${} \pm E_{\mbox{\scriptsize log}}^4 
\, \log (K/\mu)$
that is logarithmically divergent,
then one may use the relation
${} m_{\mbox{\scriptsize log}}^4 
\log (K/\mu) / (32 \pi^2)
={} E_{\mbox{\scriptsize log}}^4 
\log( K/\mu)$
to associate a logarithmic pseudomass
$ m_{\mbox{\scriptsize log}} 
={} 2 \sqrt{\pi\sqrt{2}} \, E_{\mbox{\scriptsize log}} $
with a fictitious boson if the
vacuum energy is negative
or with a fictitious fermion if the
vacuum energy is positive.
One then would generalize the third
condition (\ref{condition three}) to 
\begin{equation}
m_{\mbox{\scriptsize log}}^4
+ \sum_b m_b^4 ={} \sum_f m_f^4
\qor
\sum_b m_b^4 ={} 
m_{\mbox{\scriptsize log}}^4
+ \sum_f m_f^4.
\label{pseudo condition three}
\end{equation}
\par
A theory that has the same number
of boson modes as fermion modes
with masses that obey the fourth condition 
(\ref{condition four}) and
that has masses and pseudomasses
that obey conditions 
(\ref{pseudo condition two} and
\ref{pseudo condition three})
has a vacuum energy density
that is finite and that is entirely due
to the interactions of the theory.

\section{Inclusive theories}
\label{Inclusive theories sec}

The particles of the standard model 
including gravity
do not remotely satisfy the two
conditions 
(\ref{condition one} \&
\ref{condition two}),
let alone the third condition
(\ref{condition three})
or the fourth condition
(\ref{condition four})\@.
One may check the first condition
by counting the number of boson 
and fermion modes
in the standard model.
The graviton and the photon have 4.
The 3 heavy gauge bosons have 9.
The gluons have 16.
The Higgs has 1.
So the number of boson modes is 30.
How many fermion modes are there?
If the neutrinos are Dirac,
then they have 12.
(Incidentally, even if the low-mass neutrinos
turn out to be Majorana,
they should have heavy Majorana
brothers because before weak symmetry
breaking the neutrinos and charged leptons
formed $SU(2)$ doublets.) 
The charged leptons also have 12.
The quarks have 72 = 3\,(12 +12).
So the number of 
known fermion modes is 96\@.
The standard model
has 66 more fermion modes
than boson modes.
Presumably the particles of dark matter
and those too heavy to have been 
detected at the LHC have 66 more
boson modes than fermion modes.
Many theories
of grand unification add more
gauge bosons and scalar fields 
than fermions to the standard model.
\par
One may check the second condition 
(\ref{condition two}) by computing
the sums of the squares of the masses
of the particles of the standard model.
The sum of the squares of the 
boson masses is
$m_H^2 + 3 m_Z^2
+ 6 m_W^2 = {} 79,380.27$ GeV$^2$\@.
The sum of the squares of the 
fermion masses is
$12 (m_t^2+m_b^2 + m_c^2 + m_s^2
+ m_d^2 + m_u^2) + 
4 ( m_\tau^2 + m_\mu^2 + m_e^2)
+ 12 m_\nu^2 ={} 359,390$ GeV$^2$\@.
So the difference is 
$\sum_b m_b^2 - \sum_f m_f^2 
= {} - 280,010$ GeV$^2$\@.
Presumably the masses of the missing 
66 boson modes balance this equation.
\par
Dark matter, dark energy, the
particles of the standard model,
and all particles not yet detected
may be different aspects of a single 
inclusive theory.
The inclusive theory 
should have as many boson modes 
as fermion modes
so that the quartic divergences
cancel.
The inclusive theory
should also obey the second condition 
(\ref{condition two})  
when an appropriate 
pseudomass term representing
a possible quadratic divergence
of the vacuum energy 
due to the interactions 
is included.
The quadratic divergence 
of the vacuum energy would then vanish.
The inclusive theory also
should obey the third condition 
(\ref{condition three}) when
an appropriate 
pseudomass term representing
the logarithmic divergence
of the vacuum energy 
due to the interactions 
is included.
The vacuum energy density of the 
inclusive theory then would be finite
but much greater than the density
of dark energy  
\(\rho_{\mbox{\scriptsize de}}\)
unless the masses of the missing
modes satisfy something
like the fourth condition 
(\ref{condition four})\@.
If the inclusive theory
did obey the fourth condition
(\ref{condition four})
as well as the first three conditions 
(\ref{condition one}, \ref{condition two},
and \ref{condition three}) 
then the vacuum energy of the
inclusive theory would vanish. 
Dark energy then would be
the finite potential energy 
of the interactions
of the inclusive theory.
The particles of dark matter 
would be among those
particles that must be
added to the standard model
to make the inclusive theory 
obey the four conditions 
(\ref{condition one}, \ref{condition two},
\ref{condition three}, \ref{condition four})
when appropriate pseudomass
terms are included.

\begin{acknowledgments}
I should like to thank 
Randolph A. Reeder for
helpful email.
\end{acknowledgments}

\bibliography{physics}
 
\end{document}